\documentclass[showkeys,showpacs,superscriptaddress,twocolumn,aps,prl,reprint,10pt,nofootinbib]{revtex4-1}
\usepackage{xcolor}
\usepackage{graphicx}% Include figure files
\usepackage{dcolumn}% Align table columns on decimal point
\usepackage[symbol]{footmisc}
\usepackage{epstopdf, epsfig}
\usepackage{sidecap}
\usepackage{multirow}
\usepackage{amsmath}
\usepackage{amssymb}
\usepackage{units}

\newcommand{\angstrom}{\mbox{\normalfont\AA}}

\usepackage{bm}%bold math

\begin{document}

\preprint{APS/123-QED}

\title{Multi-Messenger Measurements of the Static Structure of Shock-Compressed Liquid Silicon at 100\,GPa}

%% Define the affiliations %%

\newcommand{\Oxford}{Department of Physics, University of Oxford, Oxford OX1 3PU, UK}

\newcommand{\LLE}{Laboratory for Laser Energetics, University of Rochester, Rochester, New York 14623, USA}

\newcommand{\UoRMechEng}{Department of Mechanical Engineering, University of Rochester, Rochester, New York 14611, USA}

\newcommand{\LLNL}{Lawrence Livermore National Laboratory, Livermore, CA 94550, USA}

\newcommand{\Reno}{Department of Physics, University of Nevada, Reno, Nevada 89557, USA}

\newcommand{\FLF}{First Light Fusion Ltd, Unit 9/10 Mead Road, Oxford Pioneer Park, Yarnton, Oxford OX5 1QU, UK}

%% Define the author list %%

\author{H. Poole}
\email{hannah.poole@physics.ox.ac.uk}
\affiliation{\Oxford}

\author{M. K. Ginnane}
\affiliation{\LLE}\affiliation{\UoRMechEng}

\author{M. Millot}
\affiliation{\LLNL}

\author{G. W. Collins}
\affiliation{\LLE}

\author{S. X. Hu}
\affiliation{\LLE}

\author{D. Polsin}
\affiliation{\LLE}

\author{R. Saha}
\affiliation{\Reno}

\author{J. Topp-Mugglestone}
\affiliation{\Oxford}

\author{T. G. White}
\affiliation{\Reno}

\author{D. A. Chapman}
\affiliation{\FLF}

\author{J. R. Rygg}
\affiliation{\LLE}\affiliation{\UoRMechEng}

\author{S. P. Regan}
\affiliation{\LLE}

\author{G. Gregori}
\affiliation{\Oxford}

\date{\today}

\begin{abstract}
Ionic structure of high pressure, high temperature fluids is a challenging theoretical problem with applications to planetary interiors and fusion capsules. Here we report a multi-messenger platform using velocimetry and \textit{in situ} angularly and spectrally resolved X-ray scattering to measure the thermodynamic conditions and ion structure factor of materials at extreme pressures. We document the pressure, density, and temperature of shocked silicon near $100\,\unit{GPa}$ with uncertainties of 6\%, 2\%, and 20\%, respectively. The measurements are sufficient to distinguish between and rule out some ion screening models.
\end{abstract}

\maketitle

With the advent of high-power lasers, a laboratory-based exploration into extreme states of matter, such as those found in planetary interiors \cite{Benuzzi14} or during asteroid impacts \cite{Sharp06}, has been realized. 
This exotic state, referred to as warm dense matter (WDM) \cite{GrazianiB}, is characterized by temperatures and pressures on the order of $1,000\,\unit{K}$ and $100\,\unit{GPa}$. 
Experimental measurement of material behavior and structure under such conditions is paramount for testing theoretical models used in the pursuit of fusion energy \cite{Craxton15, Hu18rev} and for modeling planetary phenomena \cite{Landeau22, Badro14, Soubiran17, Wijs98}, where dynamic geophysics processes are dominated by changes in solid- and liquid-state structure.

Over the last few decades, high-energy density (HED) facilities have developed the capability of generating sufficiently long-lived WDM states to deploy suites of advanced diagnostics \cite{Glenzer16, Koenig05, Wunsch11}.
Using the X-ray free electron laser (XFEL) at the Linac Coherent Light Source, high-resolution X-ray scattering measurements have successfully probed the electronic and atomic structure of high-pressure states \cite{Fletcher15, McBride19}.
However, the compression capabilities and prepared WDM volumes (critical for seeding uniform conditions) at XFELs are limited in comparison to what is readily achieved at $\unit{kJ}$ to $\unit{MJ}$-class laser facilities.
Due to the difficulty in making standard simplifying approximations at these high-pressure states, which are expected in both Jovian planet interiors \cite{Guillot99} and fusion ignition capsules \cite{Hurricane19}, equation-of-state (EOS) development \cite{Bonitz20, Hu18} requires experimental measurements.

At such laser-facilities, diagnostic access can be comparatively constrained and probing shock-compressed matter has often been limited to single diagnostics e.g. X-ray Thomson scattering (XRTS) \cite{Kraus16, Davis16, Regan12, Lee09} or X-ray diffraction (XRD) \cite{Lazicki15, Coppari22, Rygg12, Gong23} for measuring the electronic and atomic structures, or impedance matching techniques via a velocity interferometry system for any reflector (VISAR) \cite{Henderson21}.
In such experiments the conditions inferred from these diagnostics often rely on appropriate model selections or on previous measurements of reference materials. 

Initial attempts to combine scattering and velocimetry observations to infer WDM conditions were performed by Falk \textit{et al.} \cite{Falk13}, though different measurements had to be taken over multiple shots using identical targets and drive conditions.
In this work we present a novel experimental platform, where reduction of model selection biases is obtained by combining multiple diagnostics for simultaneous \textit{in situ} structure characterization.  
Reverse Monte Carlo techniques are employed to determine the structural properties of shock-compressed matter, via measurement of the static density response. For this study, silicon was chosen due to its importance in the understanding of planetary interiors \cite{Davies20, Hirose17}, for its use as a dopant to ablators in inertial confinement fusion target designs \cite{Huser18, Edwards11} and to mitigate laser-imprint effects on multi-layer targets \cite{Hu12, Fiksel12}.

The experiments were conducted at the OMEGA-EP laser facility at the Laboratory for Laser Energetics \cite{Meyerhofer10}. A $51\,\unit{\mu m}$ thick polycrystalline silicon sample was shock-compressed to $\sim 100\,\unit{GPa}$ using a single drive laser beam delivering $\sim440\,\unit{J}$ over $10\,\unit{ns}$ with a $\sim1.1\,\unit{mm}$ diameter distributed phase plate. 
The drive laser is incident on a $11\,\unit{\mu m}$ polystyrene (C$_8$H$_8$) ablator at a $19.3\unit{^{\circ}}$ angle with respect to the target normal.
The ablator was fixed to the front of the silicon sample using a thin layer of glue ($<1\,\unit{\mu m}$).
Three additional beams were tightly focused on a $12.5\,\unit{\mu m}$ thick copper backlighter with an areal size of $4\,\unit{mm^2}$, generating a $1\,\unit{ns}$ pulse of Cu He-$\mathrm{alpha}$ X-rays centered at $E\sim8.4\,\unit{keV}$ \cite{Coppari19}. The X-ray source was placed $\sim17\,\unit{mm}$ away from the silicon sample.

\begin{figure}[t]
	\centering
	\includegraphics[width=0.995\linewidth]{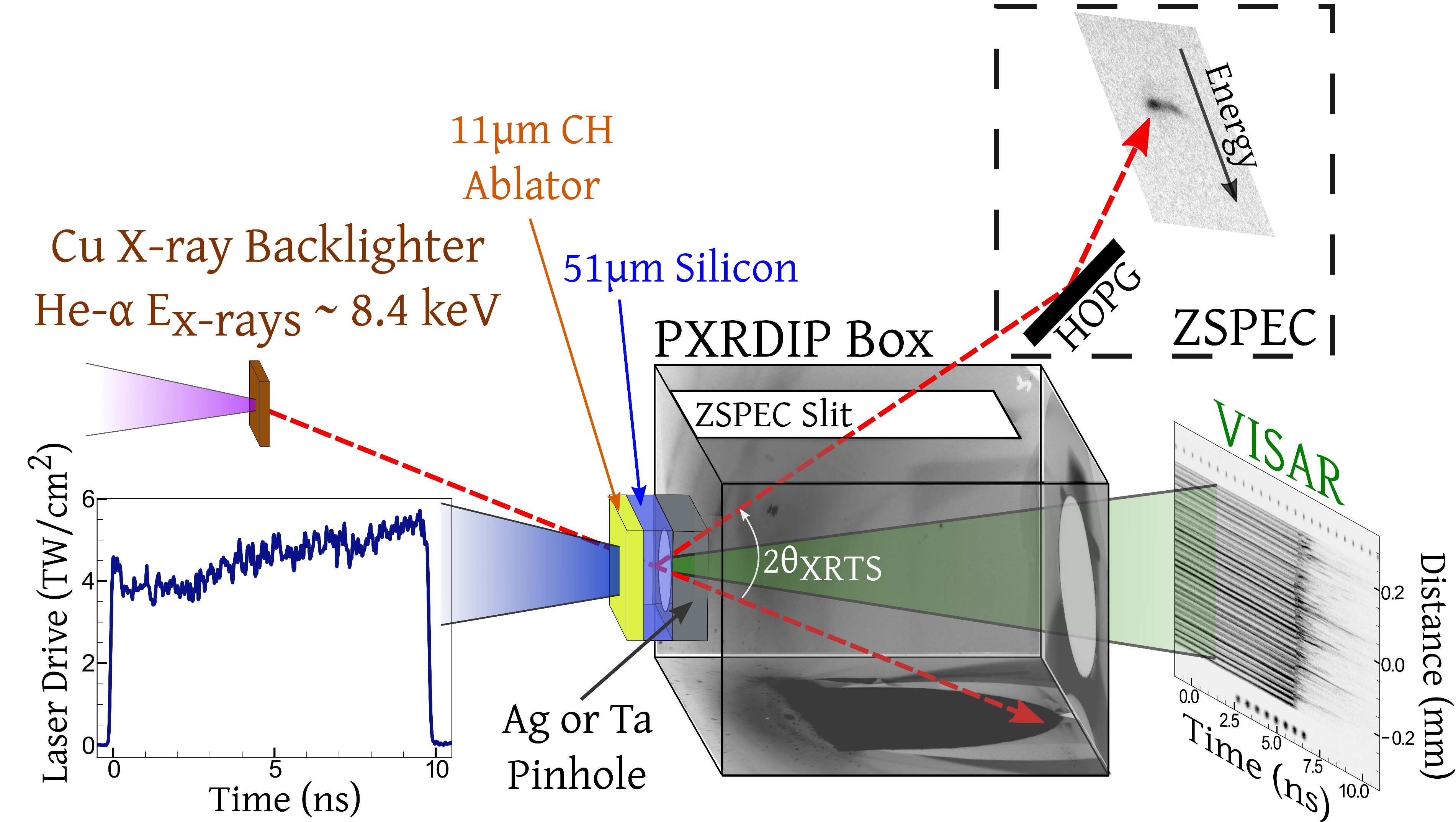}
	\caption{Experimental setup at the OMEGA-EP laser facility. The silicon target is mounted on the front of the PXRDIP box \cite{Rygg20} with a $100\,\unit{\mu m}$ thick, $0.5\,\unit{mm}$ diameter Ag or Ta pinhole. A single beam drives the CH-Si target with a tailored pulse as shown in the inset figure. The remaining three lasers generate Cu He-$\mathrm{\alpha}$ X-rays. The purple dashed lines represent the scattered X-ray paths that are collected by the XRTS and XRD IPs. The raw data shown were collected from s$30967$. NB: Not drawn to scale.
    }
	\label{fig:Experimental_setup}
\end{figure}

The experimental configuration devised to probe the structure of WDM silicon at OMEGA-EP is shown in Figure \ref{fig:Experimental_setup}. It employed a variation of the powder X-ray diffraction image plate (PXRDIP) setup \cite{Rygg12}, which uses Fujifilm BAS-MS image plates (IP's) \cite{Meadowcroft08}. 
Due to spatial constraints the X-ray diffraction only accessed momentum transfers up to $k\sim4\,\unit{\angstrom^{-1}}$ at $8.4\,\unit{keV}$.
To extend the capabilities of the PXRDIP diagnostic, a Bragg crystal zinc-spectrometer (ZSPEC) was added to measure scattering at high momentum transfer, and is capable of resolving the electronic structure of sufficiently ionized systems.
The ZSPEC consists of a $25\,\unit{mm}\times50\,\unit{mm}$ highly-oriented pyrolitic graphite (HOPG) crystal with a radius of curvature of $27\,\unit{mm}$, and placed $12.8\,\unit{cm}$ after the sample. 
As shown in the top inset in Figure \ref{fig:Experimental_setup}, the ZSPEC was fielded out of perfect von-Hamos focusing meaning the X-rays were spectrally dispersed on a curve.
The spectral analysis procedure can be found in the Supplementary Material.
The silicon sample was fitted to the front of the PXRDIP enclosure on top of a $0.5\,\unit{mm}$ diameter silver or tantalum collimating aperture pinhole, which restricts the diagnostics' line-of-sight to the central planar shock region.
These materials were chosen to ensure no fluorescence within the ZSPEC energy range, and to reduce interference between the pinhole and silicon Bragg peaks on the PXRDIP.

To measure the shock-breakout (SBO) time we fielded line-imaging VISAR which monitored the silicon sample's free surface \cite{Celliers23}. 
The streaked image inset in Figure \ref{fig:Experimental_setup} shows the SBO as a rapid disappearance of the fringes around $\sim 5\,\unit{ns}$. 
From this time we inferred the shock velocity in silicon to be $9.5\pm0.2\,\unit{km/s}$ (see Supplementary Material for details).
As silicon is opaque to the VISAR wavelength ($532\,\unit{nm}$) at the investigated conditions, a direct measurement of the silicon particle velocity could not be made, and is instead inferred from the bilinear relationship in Ref.\,\cite{Henderson21}, which for small velocities is calculated from previous high explosive measurements \cite{Pavlovskii68}. Combining this information with the Rankine-Hugoniot relations, we measured the achieved pressure-density state to be $101\pm6\,\unit{GPa}$ and $4.43\pm0.08\,\unit{g/cm^3}$.

At these conditions silicon is expected to be in the fluid state, which occurs when dynamically compressed above $30\,\unit{GPa}$ \cite{Turneaure18, McBride19}.
Whilst liquid silicon scattering, up to $30\,\unit{GPa}$, has been previously observed at XFELs \cite{McBride19}, extracting the contribution from low-Z liquids at high-power laser facilities is experimentally challenging due to limited X-ray source brightness, the presence of fluorescence, spurious scattering from the pinhole, and X-ray emission in the drive ablation plasma. 
To achieve this we quantified the contribution from the pinhole, ablation plasma and ambient sample.
The procedure is described in detail in the Supplementary Material.

\begin{figure}[!b]
	\centering
	\includegraphics[width=0.995\linewidth]{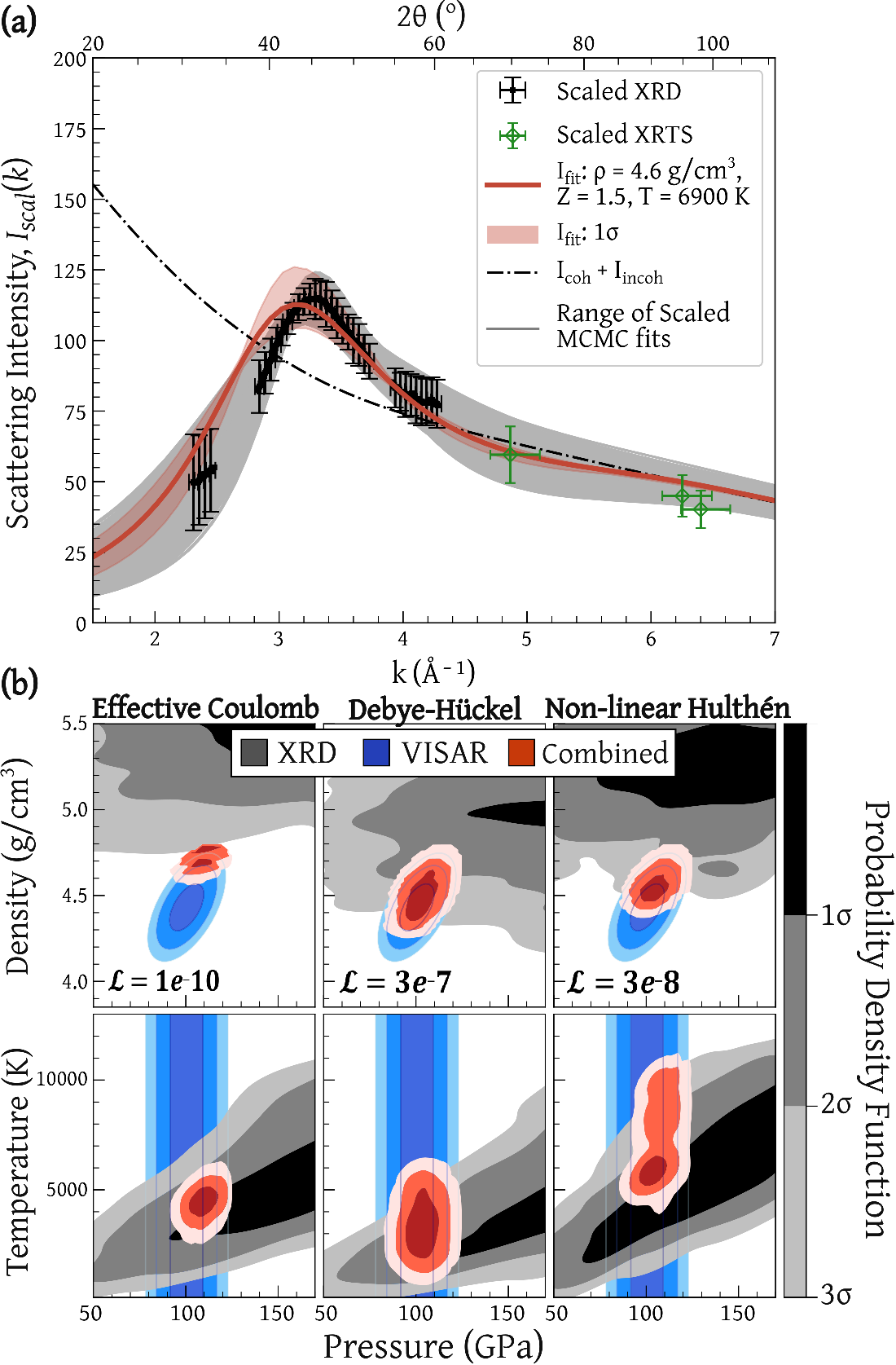}
	\caption{
 \textbf{(a)} Liquid Si diffraction signal, $I_{\mathrm{scal}}(k)$, (in blue) is shown scaled to the theoretical signal, $I_{\mathrm{fit}}(k)$, (thick red line) produced by the combined VISAR and converged MCMC conditions using the non-linear Hulth{\'e}n model. The $1\sigma$ error of $I_{\mathrm{fit}}(k)$ is shaded in red.
 The dash-dotted black line shows $I_{\mathrm{coh}}+I_{\mathrm{incoh}}$ for these values.
 The broad range of accepted MCMC fits (in gray) are scaled to the mean fit.  
 \textbf{(b)} Probability density functions in the $P$-$\rho$ and $P$-$T$ phase for VISAR (blue heat maps) and X-ray scattering (gray heat maps) analysis using each $V_{ii}$. The corresponding joint distributions are superimposed as red heat maps. In the upper grid the likelihood, as defined in equation \ref{eqn:likelihood}, of each $V_{ii}$ is shown.
        }
	\label{fig:MCSS}
\end{figure}

As shown in Figure \ref{fig:PXRDIP}(a), a broad scattering feature, attributed to liquid silicon, is observed around $2\theta\sim45\unit{^{\circ}}$.
Due to the PXRDIP's geometry and the broad band X-ray emission from the laser generated plasma plume, shadows from the box appear on the IP's, preventing a complete azimuthal integration in $\phi$-space. Instead, a partial integration is performed by selecting regions with reduced contamination from the aforementioned sources.
The resultant signal for a reference shot (s30970), which contained only the pinhole and ablator, and a driven silicon sample (s30967) are shown in Figure \ref{fig:PXRDIP}(b) in green and blue, respectively.
The final liquid silicon scattering signal, $I_{\mathrm{liq}}(k)$, shown in Figure \ref{fig:MCSS}(a) is obtained by subtracting the reference shot from the driven sample, and excluding the $2\theta$ regions around the pinhole Bragg peaks. Further details can be found in the Supplementary Material.
A $2\theta$ error of $\sim0.5\unit{^{\circ}}$ is taken to be the average deviation of the observed pinhole Bragg peaks from their expected values.

Additionally, the fraction of shocked (fluid) material within the probe volume was inferred using the ZSPEC diagnostic
by comparing data obtained with varying time delays between the drive laser and X-ray probe.
As the volume of liquid silicon increases, the elastic scattering signal recorded on the XRTS, fielded in-between Bragg peaks, becomes more intense.
From the elastic signal measured on s30967, the volume fraction was found to be $\sim0.6$ (see Supplementary Material). This gives further evidence that the diffuse signal observed on the X-ray diffraction is dominated by liquid-state silicon.

\begin{figure}[t]
	\centering
	\includegraphics[width=0.995\linewidth]{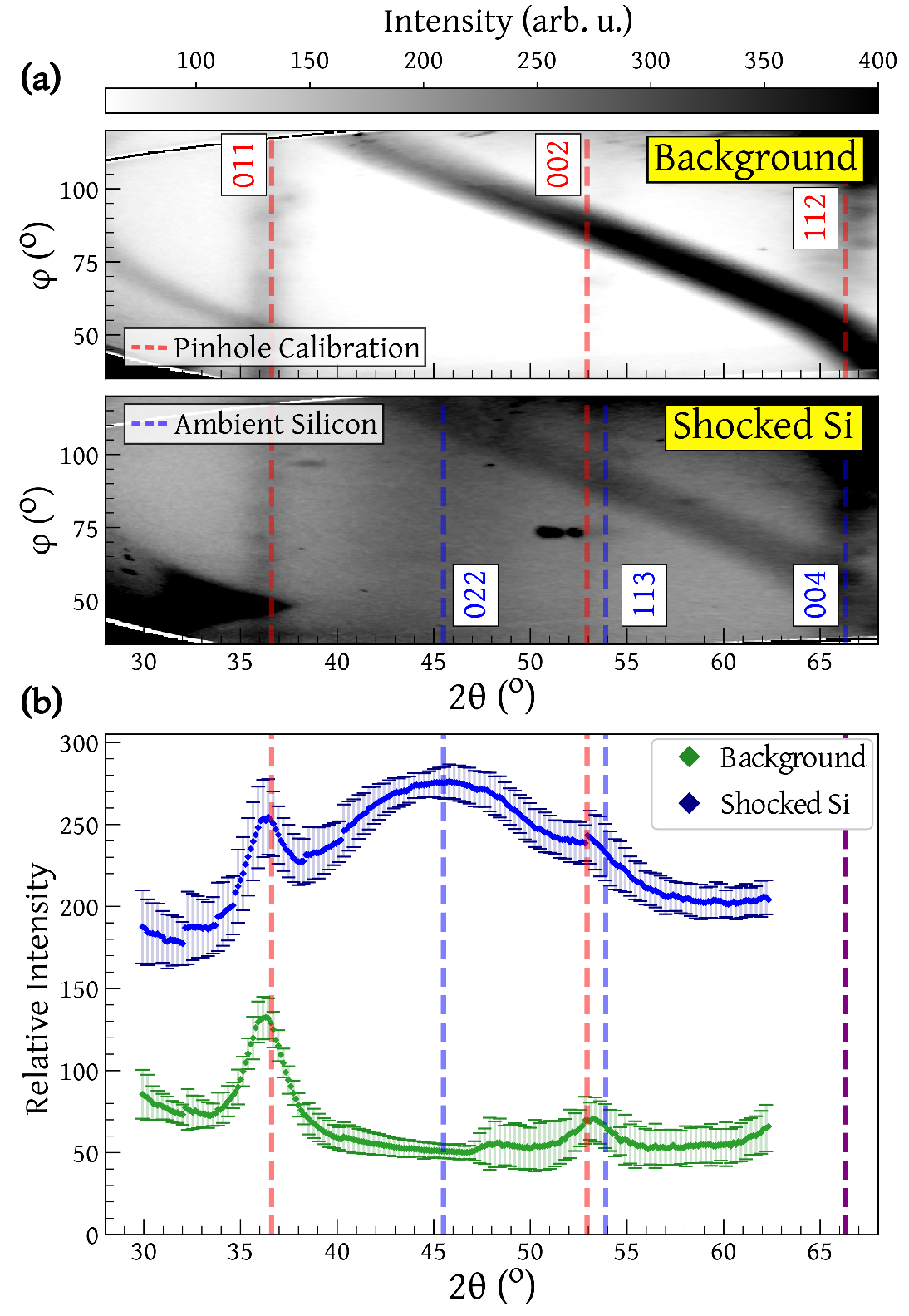}
	\caption{
        \textbf{(a)} X-ray diffraction data, projected into $2\theta$-$\phi$ space (see Supplementary Material) \cite{Rygg20}, from background shot s30970, where no Si was placed in the target holder, and the liquid Si diffraction from s30967.
        The superimposed red and blue dashed vertical lines are the expected $2\theta$ Bragg diffraction peaks of the Ta pinhole and ambient silicon, respectively.
        \textbf{(b)} Relative intensities of the partial-$\phi$ integrated scattering shown for the background (in green) and shock-compressed silicon (in blue) shots.
        }
	\label{fig:PXRDIP}
\end{figure}

At high momentum transfers the liquid scattering signal is the result of coherent, $I_{\mathrm{coh}}(k)$, incoherent, $I_{\mathrm{incoh}}(k)$, and multiple, $I_{\mathrm{m}}(k)$, scattering. As the silicon thickness is small relative to its attenuation length, $I_{\mathrm{m}}(k)$ is assumed to be negligible.  
The experimentally measured $I_{\mathrm{liq}}(k)$ is therefore related to the normalized 
ion-ion structure factor, $S_{\mathrm{ii}}(k)$, via 
\cite{Drewitt21, Singh22},
\begin{align}
    \label{eqn:structure_factor}
    \frac{I_{\mathrm{liq}}(k)}{\gamma}
    \equiv
    I_{\mathrm{scal}}(k) 
    = &\,
    I_{\mathrm{coh}}(k)\left[S_{\mathrm{ii}}(k)-1\right]
    \nonumber\\
    & +
    \left[I_{\mathrm{coh}}(k) + I_{\mathrm{incoh}}(k)\right]\,,
\end{align}
where $I_{\mathrm{coh}}(k)= \left|f(k)+q(k)\right|^2$, with $f(k)$ the form factor of the tightly bound electrons and $q(k)$ that of the free electrons that follow the ion motion \cite{Chihara00}.
The factor $\gamma$ is a scaling constant defined such that $I_{\mathrm{scal}}(k\to\infty)=I_{\mathrm{coh}}(k) + I_{\mathrm{incoh}}(k)$.
To be experimentally obtained, momentum transfers in excess of $10\,\unit{\angstrom^{-1}}$ are required, a regime not currently accessible at high-power laser facilities.
Here, $I_{\mathrm{incoh}}(k)$ is obtained using the tabulated values from Ref.\;\cite{Hubbell75} and $I_{\mathrm{coh}}(k)$ is simulated using the multi-component scattering spectra (MCSS) code \cite{MCSS}.
As detailed further in the Supplementary Material, $\gamma$ is left proportional to a free random Gaussian scalar with a standard deviation equal to the noise of the raw data.

The large parameter space, $\Psi(\rho, T, Z)$, is explored using a Markov-Chain Monte Carlo (MCMC) procedure \cite{Kasim19, Poole22}. This uses Bayesian inference to determine the likelihood of a set of parameters producing the experimental spectrum based on an acceptance percentage $P\left[I_{\mathrm{scal}}(k)|\Psi\right]=\mathrm{e}^{-\beta_{\mathrm{cost}}}$ with
\begin{equation}
    \label{eqn:MCMC_cost}
    \beta_{\mathrm{cost}}
    =
    \mathrm{max}
    \left[\frac{I_{\mathrm{fit}}(k) - I_{\mathrm{scal}}(k)}{\sqrt{2}\sigma\Sigma}\right]^2\,,
\end{equation}
where $\Sigma$ is the error on $I_{\mathrm{scal}}$, and $\sigma=0.5$ is a scalar chosen to allow acceptance freedom within data uncertainty. The investigated parameter space assumed a uniform distribution with linear sampling for the density, $2.33\leq \rho\,(\mathrm{g/cm^3})\leq 6$, ionization, $0\leq Z \leq 14$, and temperature, $10^3\leq T_e=T_i\,(\mathrm{K})\leq 1.1\times10^4$. 

\begin{figure}[b]
	\centering
	\includegraphics[width=0.995\linewidth]{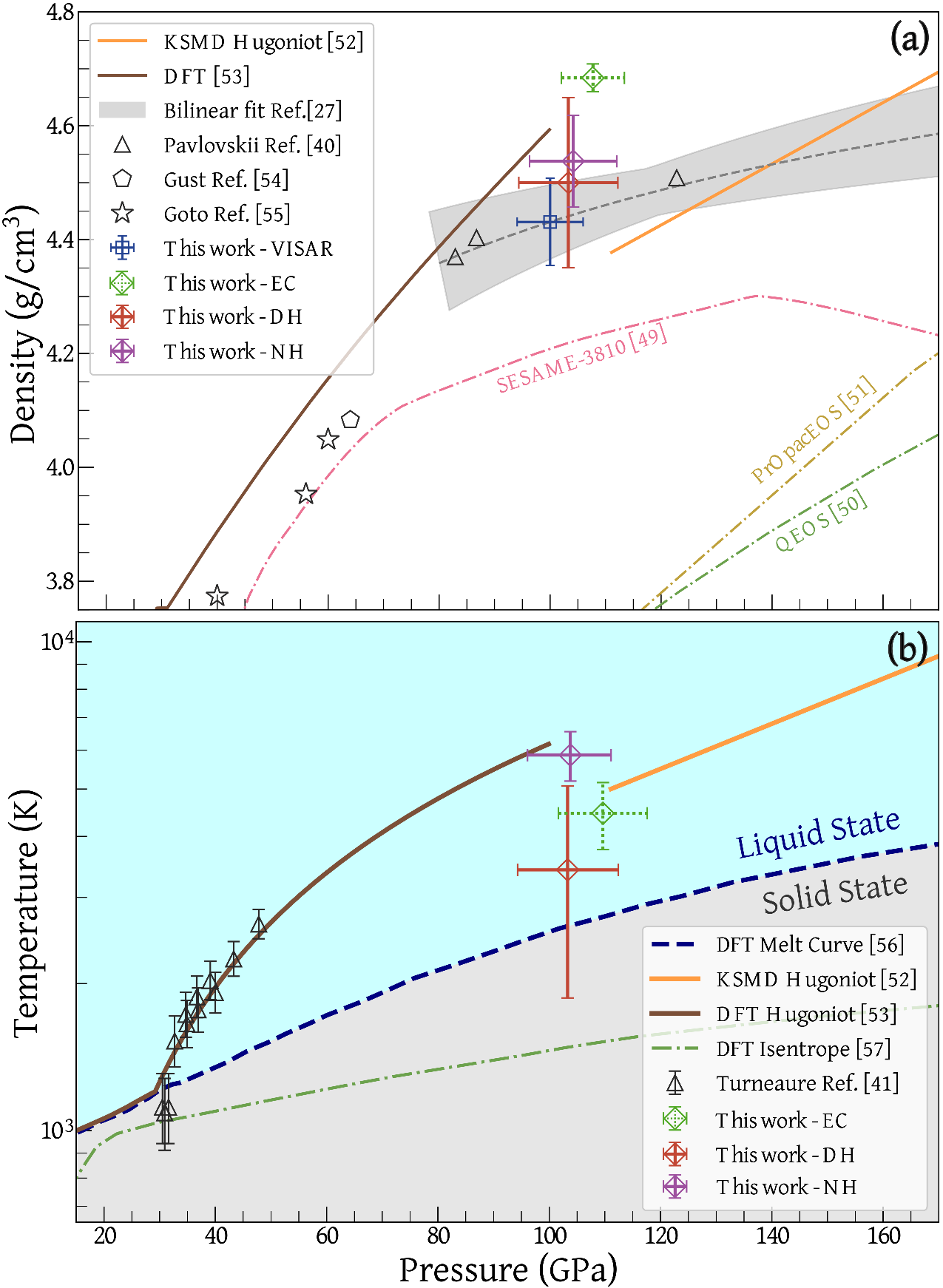}
	\caption{
	\textbf{(a)} The principal silicon Hugoniot where this work is compared to SESAME-3810, \cite{SESAME}, quotidian equation-of-state (QEOS) \cite{QEOS}, PrOpacEOS \cite{HELIOS}, $ab\,\,initio$ Kohn-Sham DFT molecular-dynamics (KSMD) \cite{Hu16}, principle Hugoniot from DFT \cite{Paul19DFT}, and previous experimental work collected via conservation methods \cite{Pavlovskii68, Gust71, Goto82}. The bilinear fit \cite{Henderson21} used to infer particle velocity is shown  as a filled gray bar.
	\textbf{(b)} The silicon pressure-temperature phase diagram comparing the combined $1\sigma$ error for each $V_{ii}$ to the measured and predicted melt curve \cite{Paul19}, the DFT isentrope \cite{Paul22} and previous shocked silicon experiments \cite{Turneaure18} where the temperature was inferred using molecular dynamics \cite{Strickson2016}.
        }
	\label{fig:HUG}
\end{figure}

Simulating $S_{\mathrm{ii}}(k)$, however, is subject to model biases and requires appropriate selection of electron and ion interactions. 
Measurement of the liquid structure factor opens the opportunity for direct model comparison.
In the partially ionized, low density state, the ion-ion interaction potential, $V_{ii}(k)$ is commonly modeled using Debye-H\"uckel (DH) \cite{DH}. This work compares the DH model with the bare (unscreened) effective Coulomb (EC) interaction and a model non-linear Hulth\'en (NH)  interaction \cite{NH}; the latter approximately describes screening beyond the DH approach. For the screening cloud, $q(k)$, large momentum transfers in high-density matter have shown deviation from the simple DH model as a result of finite-wavelength screening (FWS) \cite{Chapman15}. As detailed in the Supplementary Material, the simulated liquid scattering is comparatively insensitive to each $q(k)$ model and FWS was chosen for the MCMC analysis.

In Figure \ref{fig:MCSS}(a) the range of accepted fits after MCMC convergence using the NH model are shown in gray. 
The signal from the XRTS recorded on shots that were probed after shock breakout (where the liquid volume fraction $> 0.9$) are compared in green against the angularly resolved scattering in Figure \ref{fig:MCSS}(a), extending the effective $k$ range. They show good agreement with the MCMC results.

Using a suitable theoretical description, the total plasma pressure can be determined from the range of accepted fits.
Under conditions of strongly coupled ions and degenerate electrons, in which screening is expected to be significant, a reasonable framework is the `two-fluid' model discussed by Vorberger \textit{et al.}\;\cite{Vorberger13, Ebeling20} (see Supplementary Material). 
The converged probability density functions $\mathrm{Pr}(P, \rho)$ and $\mathrm{Pr}(P, T)$, for each ion-ion potential model, are shown in gray in Figure \ref{fig:MCSS}(b) and compared, in blue, to the $P$-$\rho$ state inferred using VISAR. 
We can combine these concurrent diagnostics to find joint $P$-$\rho$ probability density functions which are superimposed in Figure \ref{fig:MCSS}(b) as red heat maps.

The likelihood of each ion-ion potential model given the VISAR information is defined as the sum of its joint probability distribution,
\begin{equation}
    \label{eqn:likelihood}
    \mathcal{L}\left(V_{ii}|\mathrm{VISAR}\right)
    =
    \sum_{\rho, P} \mathrm{Pr}_m(P, \rho)\times\mathrm{Pr}_v(P, \rho)\,,
\end{equation}
where $m$ and $v$ denote the MCMC and VISAR probability density functions, respectively.
These likelihoods are indicated in the upper grid of Figure \ref{fig:MCSS}(b). They show that comparatively, the effective-Coulomb model is a poor representation of the liquid silicon state. This is expected as it does not account for screening effects.

Unlike the VISAR diagnostic, the MCMC convergence of the X-ray scattering analysis is dependent not only on pressure and density, but also on temperature. The combined $\mathrm{Pr}(P, \rho)$ can therefore be propagated into temperature space. This re-distributes the X-ray scattering $\mathrm{Pr}(P, \rho, T)$ to penalize where the density and pressure disagree with VISAR.
Further details of this process can be found in the Supplementary Material.
The resultant $\mathrm{Pr}(P, \rho, T)$ are used to find the combined $1\sigma$ errors in the pressure-temperature phase, shown in red in the lower grid of Figure \ref{fig:MCSS}(b). 
The simulated X-ray diffraction fits, $I_{\mathrm{fit}}$, produced by the conditions inferred when combining VISAR and the NH MCMC convergence are shown in red in Figure \ref{fig:MCSS}(a).

In Figure \ref{fig:HUG} the VISAR and MCMC combined $1\sigma$ $P$-$\rho$ and $P$-$T$ for each ion-ion potential model are plotted on the principal Hugoniot and compared to density functional theory (DFT) calculations \cite{Hu16, Paul19DFT}, the Si melt \cite{Paul19}, and previous experimental work.
Despite having the closest agreement with VISAR in $P$-$\rho$, the temperature predicted by the commonly used Debye-H{\"u}ckel model falls below the Hugoniot state.
Instead we find the implementation of a Hulth{\'e}n potential \cite{NH}, which estimates non-linear screening regimes beyond DH, better describes the thermodynamic conditions.

This report demonstrates how a novel experimental platform at a high-power laser-facility collects detailed information on the structure of shock-compressed matter at HED conditions. 
Whilst previous work on liquid silicon structure has been limited to pressures of $\sim50\,\unit{GPa}$ \cite{McBride19, Turneaure18}, this platform is highly scalable to different pressures and materials.
Our presented analytical technique applied Markov-Chain Monte Carlo analysis to the observed liquid diffraction signal and combined it with the concurrent VISAR $P$-$\rho$ measurement.
This provided $1\sigma$ uncertainties on the shock-compressed state that are equivalent to previous experimental work without relying on EOS models.
In addition, this platform enables the investigation of ion-ion interaction potentials. We found that accounting for screening beyond the linear Debye-H{\"u}ckel approach via a Hulth{\'e}n potential was required for agreement between the measured liquid silicon state and Hugoniot predictions.
This platform therefore facilitates both mapping solid-to-liquid transitions for high pressure states and reducing model selection biases.

%% Acknowledgements
\begin{acknowledgments}
This material is based upon work supported by the Department of Energy National Nuclear Security Administration under Award Number DE-NA0003856, the University of Rochester, and the New York State Energy Research and Development Authority and the National Science Foundation under Grant No. PHY-2045718.
Part of this work was prepared by LLNL under Contract No. DE-AC52–07NA27344. The PXRDIP data was analyzed with LLNL AnalyzePXRDIP package. 

This report was prepared as an account of work sponsored by an agency of the U.S. Government. Neither the U.S. Government nor any agency thereof, nor any of their employees, makes any warranty, express or implied, or assumes any legal liability or responsibility for the accuracy, completeness, or usefulness of any information, apparatus, product, or process disclosed, or represents that its use would not infringe privately owned rights. Reference herein to any specific commercial product, process, or service by trade name, trademark, manufacturer, or otherwise does not necessarily constitute or imply its endorsement, recommendation, or favoring by the U.S. Government or any agency thereof. The views and opinions of authors expressed herein do not necessarily state or reflect those of the U.S. Government or any agency thereof.
\end{acknowledgments}

%% Construct bibliograhy %
\bibliographystyle{apsrev}
\bibliography{main}

\end{document}

\preprint{APS/123-QED}

\title{Supplementary Material for \\ ``Multi-Messenger Measurements of the Static Structure of Shock-Compressed Liquid Silicon at 100\,GPa''}

%% Define the affiliations %%

\newcommand{\Oxford}{Department of Physics, University of Oxford, Oxford OX1 3PU, UK}

\newcommand{\LLE}{Laboratory for Laser Energetics, University of Rochester, Rochester, New York 14623, USA}

\newcommand{\UoRMechEng}{Department of Mechanical Engineering, University of Rochester, Rochester, New York 14611, USA}

\newcommand{\LLNL}{Lawrence Livermore National Laboratory, Livermore, CA 94550, USA}

\newcommand{\Reno}{Department of Physics, University of Nevada, Reno, Nevada 89557, USA}

\newcommand{\FLF}{First Light Fusion Ltd, Unit 9/10 Mead Road, Oxford Pioneer Park, Yarnton, Oxford OX5 1QU, UK}

%% Define the author list %%

\author{H. Poole}
\email{hannah.poole@physics.ox.ac.uk}
\affiliation{\Oxford}

\author{M. K. Ginnane}
\affiliation{\LLE}\affiliation{\UoRMechEng}

\author{M. Millot}
\affiliation{\LLNL}

\author{G. W. Collins}
\affiliation{\LLE}

\author{S. X. Hu}
\affiliation{\LLE}

\author{D. Polsin}
\affiliation{\LLE}

\author{R. Saha}
\affiliation{\Reno}

\author{J. Topp-Mugglestone}
\affiliation{\Oxford}

\author{T. G. White}
\affiliation{\Reno}

\author{D. A. Chapman}
\affiliation{\FLF}

\author{J. R. Rygg}
\affiliation{\LLE}\affiliation{\UoRMechEng}

\author{S. P. Regan}
\affiliation{\LLE}

\author{G. Gregori}
\affiliation{\Oxford}

\date{\today}
\maketitle

\section{Spectrally resolved X-ray scattering}{\label{sec:XRTS}}

\begin{table}[b]
    \centering
    \renewcommand{\arraystretch}{1.5}
    \caption{\label{SItab:EXP} Experimental parameters for all shots including the total incident energy of the shock-compression drive, $E_{\mathrm{drive}}$, and X-ray, $E_{\mathrm{xray}}$, lasers.}
    \begin{ruledtabular}
    \begin{tabular}{l|cccccccccc}
    \textbf{Shot} & $2\theta_{\mathrm{XRTS}}\,\,(\si{^{\circ}})$ & Pinhole & $h_{\mathrm{CH}}\,\si{(\mu m)}$& $h_{\mathrm{Si}}\,\si{(\mu m)}$ &$E_{\mathrm{drive}}\,(\si{J})$ & $E_{\mathrm{xray}}\,(\si{J})$ & $t_{\mathrm{drive}}\,(\si{ns})$ & $t_{\mathrm{xray}}\,(\si{ns})$ & $t_{\mathrm{SBO}}\,(\si{ns})$ & $L_{\mathrm{coh}}$  \\ \hline
    \multicolumn{11}{c}{\textbf{Background target - No Silicon}}   \\ \hline
    $30970$     & $70$  & Ta        & $11\pm2$  & -         & -         & $3760.7$  & -         & -     & -    & $260\pm9$         \\ \hline
    \multicolumn{11}{c}{\textbf{Ambient conditions}}   \\ \hline
    % $30965$     & $95$  & Ta        & $11\pm2$  & $51\pm1$  & -         & $3772.7$  & -         & -     & -             \\
    $30968$     & $70$  & Ta        & $11\pm2$  & $51\pm1$  & -         & $3618.6$  & -         & -     & -      & $430\pm10$       \\ \hline
    % \multicolumn{10}{c}{\textbf{Shock-compressed with LiF witness}}   \\ \hline
    % $30969$     & -     & -         & $11\pm2$  & $51\pm1$  & $428.4$   & -         & $-0.033\pm0.025$  & -     & $7.55\pm0.05$     \\ \hline
    \multicolumn{11}{c}{\textbf{Shock-compressed conditions}}   \\ \hline
    $30964$     & $95$  & Ta        & $11\pm2$  & $51\pm1$  & $441.3$   & $3707.5$  & $-0.049\pm0.025$  & $5$   & $5.35\pm0.04$  & $615\pm10$   \\
    $30967$     & $70$  & Ta        & $11\pm2$  & $51\pm1$  & $429.0$   & $3908.4$  & $-0.15\pm0.025$   & $4.8$ & $6.22\pm0.06$  & $580\pm10$   \\
    $33538$     & $70$  & Ag        & $11\pm2$  & $51\pm1$  & $437.9$   & $3778.3$  & $-0.002\pm0.025$  & $5$   & $5.20\pm0.04$  & $875\pm20$   \\
    $33541$     & $98$  & Ag        & $11\pm2$  & $51\pm1$  & $422.5$   & $3835.0$  & $0.014\pm0.025$   & $5$   & $4.91\pm0.03$  & $534\pm8$   \\

    \end{tabular}
    \end{ruledtabular}
\end{table}

\begin{figure}[t]
	\centering
	\includegraphics[width=0.995\textwidth]{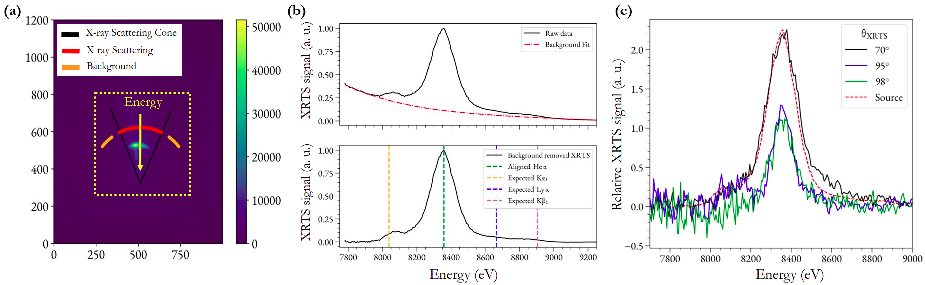}
	\caption{
        \textbf{(a)} Raw intensity of calibration s33544 where the Cu foil was placed in the target holder of the PXRDIP box. 
        \textbf{(b)} In the top plot the XRTS spectrum extracted after integrating along each energy arc is shown. A further polynomial background fit (red dash-dotted line) is subtracted from the overall signal to produce the spectrum in the lower plot. The energy axis is calibrated to the Cu He-alpha peak. The expected positions of the remaining Cu transitions are shown as vertical dashed lines.
        \textbf{(c)} Comparison of the spectrally resolved XRTS signal for the post-SBO shots where $V_\mathrm{l}/V_\mathrm{s} > 0.9$. The source function from (b) is scaled to the $\theta_{\mathrm{XRTS}}=70\si{^{\circ}}$ scattering signal and is shown as a red dashed line.
        }
	\label{SIfig:XRTS}
\end{figure}

The zinc-spetrometer (ZSPEC) diagnostic used to record the X-ray Thomson scattering consisted of a highly-oriented pyrolitic graphite (HOPG) crystal placed $12.8\,\unit{cm}$ after the sample. 
The central Bragg angle was $13.2^{\circ}$ with an angular spread of $2.5^{\circ}$.
Its image plate (IP) was protected by a $5\,\unit{mm}$ Be filter and mounted $13.1\,\unit{cm}$ from the crystal meaning the ZSPEC was fielded out of perfect von-Hamos focusing. 
This resulted in the spectrally resolved X-rays being scattered on to a cone which can allow for some spatial information to be resolved. In the raw intensity image shown in Figure \ref{SIfig:XRTS}(a), the X-ray scattering cone is highlighted in black and the scattered photon energy increases to its point. 
The dispersion of the ZSPEC is \cite{Pak04}
${\Delta E}/{\Delta x} \sim 7\,\unit{eV/pixel}$ where the pixels are $50\,\unit{\mu m}$ wide.
To extract the spectrally resolved spectrum, the total scattering signal along each energy arc is determined. The pixels falling along an arc, such as those highlighted in red, have an average pixel background value (determined from the orange arcs which fall outside of the X-ray scattering cone) removed, and are then summed. 
This gives an integrated scattering signal, $I(x)$, for a given spatial position $x$ on the IP. The spectrum is converted from spatial to energy space using,
\begin{equation}
    \label{eqn:XRTSdis}
    I(E)
    =
    \mbox{\Large\( %
    \frac{hc}{2d\,\,\mathrm{sin}\left(\mathrm{tan}^{-1}\left(
    \frac{I(x) + D_{c}\mathrm{sin}\theta_{0} + D_{ip}\mathrm{sin}\theta_{0}}{D_{ip}\mathrm{cos}\theta_{0} + D_{c}\mathrm{cos}\theta_{0}}
    \right)\right)}
    \)}
    \,,
\end{equation}
where $2d = 0.67\,\si{nm}$ is the HOPG lattic spacing, $D_c=12.8\,\si{cm}$ is the distance from the source to the HOPG crystal, $D_{ip}=13.1\,\si{cm}$ is the distance from the HOPG crystal to the IP, and $\theta_0=13.2\si{^{\circ}}$ is the central Bragg angle on the crystal. 
In the upper plot of Figure \ref{SIfig:XRTS}(b), the XRTS spectrum from the Cu X-ray source is shown. After a further polynomial fit to the background is removed, the total background removed spectrally resolved XRTS spectrum is given in the lower plot of Figure \ref{SIfig:XRTS}(b).
The energy calibration is performed by aligning the peak signal with the Cu Helium-alpha X-rays at $8.358\,\si{keV}$. It can be seen in the lower plot that the remaining expected Cu transitions align with the lower intensity peaks on the spectrum.

A comparison of the XRTS spectra produced by the data shots taken after SBO (i.e. where $V_\mathrm{l}/V_\mathrm{s} > 0.9$) are shown in Figure \ref{SIfig:XRTS}(c) against the source function. As expected with low silicon ionization, there is no resolvable inelastic scattering signal above the source function. 
The XRTS signals can therefore not be used independently to extract the plasma parameters, but can be combined with the X-ray diffraction.

\begin{figure}[t]
	\centering
	\includegraphics[width=0.75\textwidth]{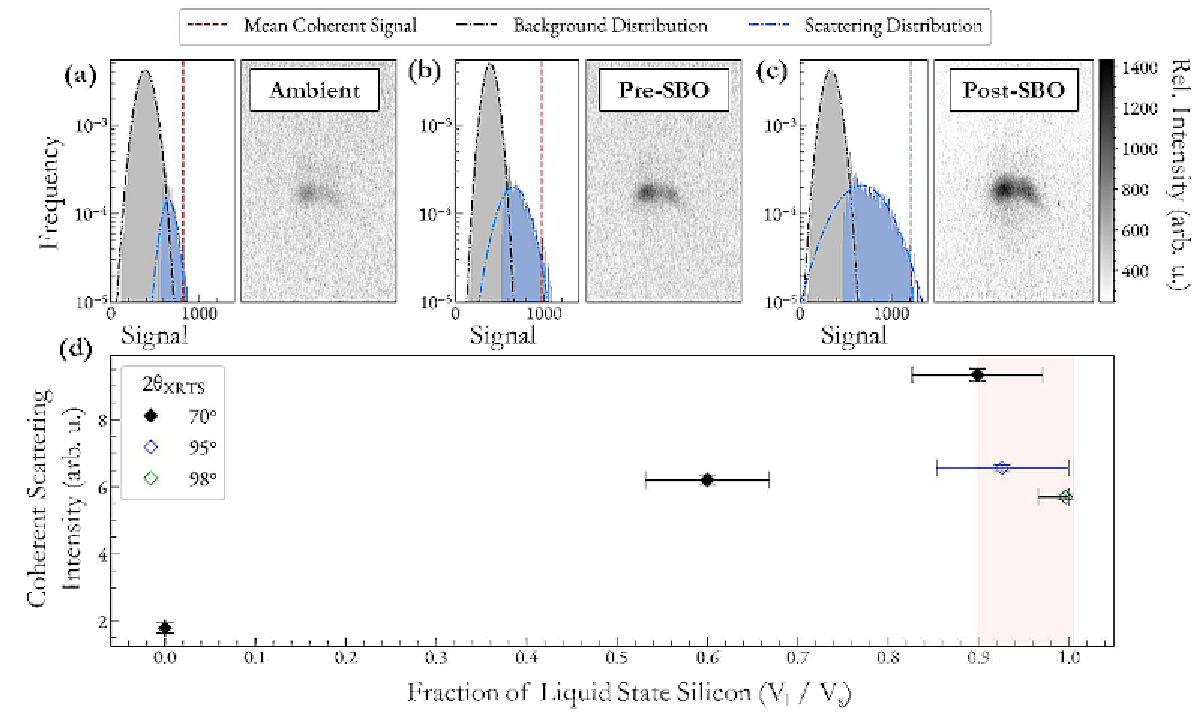}
	\caption{
        The XRTS signal intensities focused around the coherent scattering arc for ambient s30968 \textbf{(a)}, pre-SBO s30967 \textbf{(b)} and post-SBO s33538 \textbf{(c)} at $\theta_{\mathrm{XRTS}}=70\si{^{\circ}}$. The raw scattering image is shown on the right of each plot and their corresponding scattering signal histograms are shown on the left. 
        % Each histogram in gray is the total range of observed signal intensity with the fitted black dash-dotted Gaussian distributions being dominated by the background signal. 
        % The overlaid blue histograms isolate the scattering signal after the background Gaussian is removed. 
        The red dashed lines are the mean coherent signal intensities, $L_{\mathrm{coh}}$.
        \textbf{(d)} Relative intensity of elastic XRTS signal ($\propto S_{ii}$) against the fraction of liquid silicon, $V_{\mathrm{l}}/V_{\mathrm{s}}$, for all $2\theta_{\mathrm{XRTS}}$. Highlighted in red are the shots taken after shock breakout. For these shots the liquid fraction is greater than $\sim0.9$ and the scattering is assumed to be only from liquid silicon.
        The liquid fraction of the unfilled diamond points at $95\si{^{\circ}}$ and $98\si{^{\circ}}$ are determined from HELIOS simulations due to insufficient ambient data at these scattering locations.
        }
	\label{SIfig:COH}
\end{figure}

To compare the XRTS to the angularly resolved X-ray diffraction signal, the XRTS must be spectrally integrated to give $I(k) = \sum_{\omega}I(k, \omega)$. However, as only the coherent scattering signal is clearly observed over the background noise, an integration over all $\omega$ cannot be performed. Taking the peak intensity of the spectrally resolved signals shown in Figure \ref{SIfig:XRTS}(c) increases the measurement uncertainty due to the compounded error of isolating the X-ray scattering cone and from the integration methods performed over each energy arc.
Instead, as demonstrated in Figure \ref{SIfig:COH}, the relative coherent scattering signals, $L_{\mathrm{coh}}$, are determined by removing fitted background Gaussian distributions from the overall signal intensities - isolating the target scattering. 
The signals are corrected for filtering, polarization, backlighter distance and relative thickness of the silicon seen by the X-ray source.
The subsequent Gaussian fits to these scattering histograms yields, $L_{\mathrm{coh}} = (\mu_{s} + 2\,\sigma_{s}) - \mu_b$, where $s$ and $b$ denote the Gaussian distributions to the scattering and background signals, respectively. 

For the ambient s30968, where $2\theta_{\mathrm{XRTS}}=70\si{^{\circ}}$, the Ta Bragg peak from the (112) lattice plane is resolved by the XRTS. This pinhole scattering signal is isolated from the silicon scattering by subtracting the XRTS coherent signal from the reference s30970, where only the Ta pinhole was loaded in the target holder. The coherent scattering intensities are listed in Table \ref{SItab:EXP}.

As discussed in the main paper, comparing the coherent scattering intensities of ambient and shocked silicon provided information on the fraction of shocked (fluid) silicon within the probe volume.
Using a simple scattering model as described by Pelka \textit{et al.}\,\cite{Pelka10}, which is based on the approach of Chihara \cite{Chihara00}, the time-averaged volume fraction of liquid ($l$) to solid ($s$) silicon present during the scattering event is calculated as,
\begin{equation}
    \label{eqn:VlVtot}
    \frac{V_l}{V_{\mathrm{s}}}
    =
    \frac{L_{\mathrm{coh}}^{l}}{L_{\mathrm{coh}}^{s}}
    \,
    \frac{S_{\mathrm{tot}}^{s}}{S_{\mathrm{tot}}^{l}}
    =
    \frac{L_{\mathrm{coh}}^{l}}{L_{\mathrm{coh}}^{s}}
    \,
    \frac{Z_{Si}\left[1-I_{\mathrm{coh}}(k)/Z_{Si}^2\right]}{I_{\mathrm{coh}}(k)S_{ii}(k)}
    \,,
\end{equation}
where $S_{\mathrm{tot}}$ are the static structure factors and $Z_{\mathrm{Si}}=14$ is the nuclear charge.
As shown in Figure \ref{SIfig:COH}, the volume fraction for the shock-compressed silicon states was found to be $V_l/V_s>0.6$.

As shown in Figure 3(a) of the main paper, the post-shock breakout XRTS data shots (s30964, s33538 and s33541) -- where the fraction of liquid silicon is greater than $\sim0.9$ -- are compared to the diffuse angularly resolved scattering recorded on s30967. 
These $L_{\mathrm{coh}}$ signals are scaled by fitting the value at $2\theta_{\mathrm{XRTS}}=70^{\circ}$ to the range of accepted MCMC fits. As only the coherent contribution is resolved using ZSPEC, this fitting procedure uses the MCMC fits, $I_{\mathrm{fit}}$, prior to adding the incoherent scattering, $I_{\mathrm{incoh}}$. 
% As $L_{\mathrm{coh}}$ does not contain any contribution from inelastic scattering, the incoherent contribution must be first removed from the PXRDIP data.
The higher $2\theta_{\mathrm{XRTS}}$ signals are then determined by their scattering intensity relative to the $70^{\circ}$ data. The total signal errors are compounded by their respective coherent scattering uncertainties and the scaling error of $L_{\mathrm{coh}}$ at $70^{\circ}$.

\section{Angularly resolved X-ray scattering}{\label{sec:XRD}}

\begin{figure}[t]
	\centering
	\includegraphics[width=0.9\textwidth]{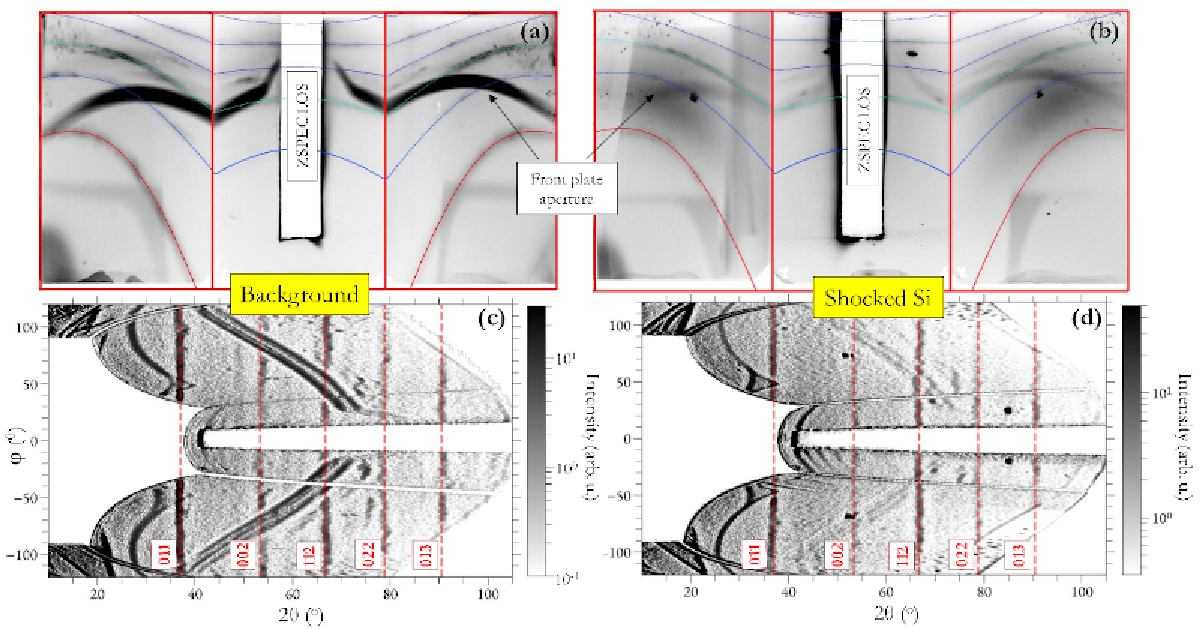}
	\caption{
        \textbf{(a)} and \textbf{(b)}: Raw images plates for background s30970 and shocked silicon s30967, respectively. The lines shown are the Bragg diffraction peaks of the Ta pinhole which are used to calibrate the geometry of the PXRDIP box. Their corresponding warped $2\theta$-$\phi$ signals at the pinhole position are shown in \textbf{(c)} and \textbf{(d)}, respectively. The scattering distributions for the pinhole used a SNIP background removal process \cite{Rygg20} which enhances sharp peaks, improving signal-to-background ratios.
        % after performing a SNIP background removal \cite{Rygg20} at the pinhole position are shown in \textbf{(c)} and \textbf{(d)}, respectively.
        % The superimposed red dashed horizontal lines are the
        % $2\theta$ Bragg diffraction peaks of the Ta pinhole which are used to calibrate and resolve the geometry of the PXRDIP box.
        }
	\label{SIfig:XRD_raw}
\end{figure}

\begin{figure}[t!]
	\centering
	\includegraphics[width=0.9\textwidth]{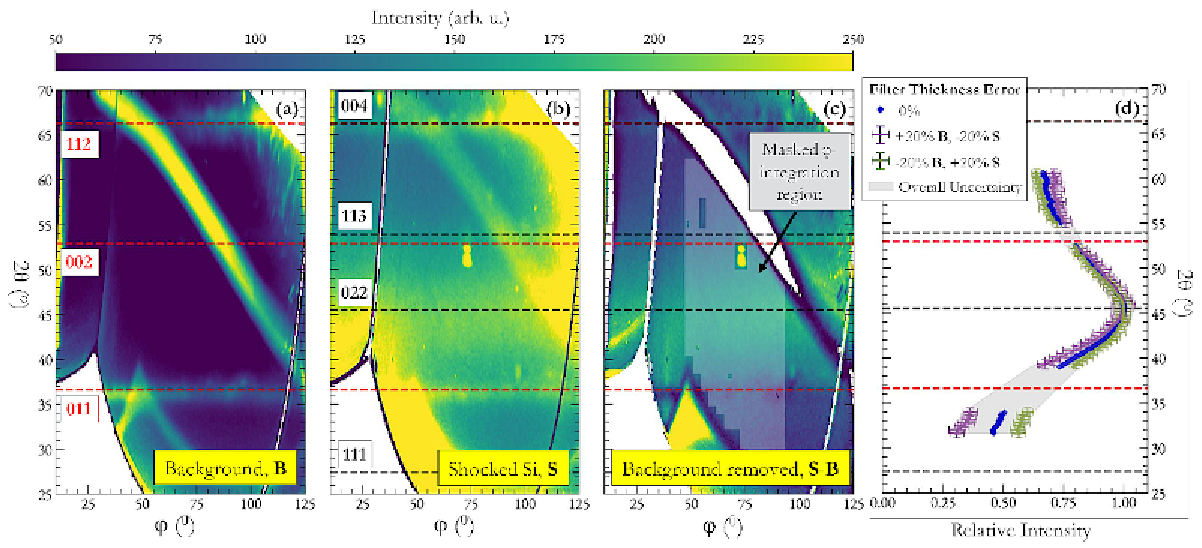}
	\caption{
        \textbf{(a)} and \textbf{(b)}: The warped, intensity corrected signals at the sample position for background s30970 and shocked silicon s30967, respectively. 
        The superimposed red and black dashed horizontal lines are the calibrated $2\theta$ Bragg diffraction peaks of the Ta pinhole and the expected ambient silicon peaks, respectively.
        \textbf{(c)} Shocked Si scattering after background removal. Artifacts from this removal process are seen at the edges of the image plates. The region selected for $\phi$-integration is highlighted in gray.
        \textbf{(d)} Shown in blue, the partial-$\phi$ integration of (c) to obtain the liquid scattering signal in $2\theta$. The purple and green lines show the effect filter thickness uncertainties have on the inferred liquid shape. The overall signal uncertainty is taken over the gray shaded region.
        }
	\label{SIfig:XRD}
\end{figure}

Using Lawrence Livermore National Laboratory's AnalyzePXRDIP procedure \cite{Rygg20}, the raw X-ray diffraction IP's, shown in Figures \ref{SIfig:XRD_raw}(a) and (b), are warped into $2\theta$-$\phi$ by using the Ta pinhole Bragg peaks as calibrants. The pinhole calibration for each shot is shown in (c) and (d).
To isolate the liquid silicon scattering the background signal must be removed from the shocked data.
Whilst a variant of the Statistics-sensitive Non-linear Iterative Peak-clipping (SNIP) algorithm is often used to isolate signal from the background in PXRDIP scattering data \cite{Rygg20}, this process is only appropriate when dealing with sharp Bragg peaks, such as shown in Figure \ref{SIfig:XRD_raw}.
As discussed in the main paper, to quantify the background signal we instead performed a series of shots to isolate each contributor.
% To achieve this we quantified the contribution from the pinhole, ablation plasma and ambient sample.
In s30966, we investigated the signal intensity recorded on the IP's when only the drive laser was present (i.e. no X-ray probe). At the investigated pressure, this was found to be negligible in comparison to the data collected with X-ray scattering events.
Comparison of the scattering recorded on the PXRDIP using an ambient silicon sample (e.g. s30968) versus for a reference s30970, where the silicon sample was removed, showed the dominant background scattering contributor to be the pinhole.
Subsequent analysis therefore uses s30970 as the background reference.

The scattering for the background s30970 and shocked s30967 shown in Figure \ref{SIfig:XRD} (a) and (b), respectively, are obtained by accounting for the filtering ($12.5\,\unit{\mu m}$ Cu and $25\,\unit{\mu m}$ Kapton), incident solid angle, polarization ($(1+\cos^22\theta)/2$) and the attenuation of the scattered X-rays.
The background subtracted shocked signal is shown in Figure \ref{SIfig:XRD}(c). Contamination at the edge of the IPs means a full integration in $\phi$-space cannot be performed. The partial-$\phi$ integration region, highlighted in gray, therefore focuses on the IP center, and excludes the $2\theta$ regions around the Ta Bragg peaks.
The resultant liquid scattering signal is shown in blue in Figure \ref{SIfig:XRD}(d).
The $12.5\,\unit{\mu m}$ Cu and $25\,\unit{\mu m}$ Kapton filters used for the PXRDIP IPs have a $20\%$ thickness uncertainty. Propagating this error through the background removal process results in the purple and olive plots shown in (d). The intensity error of each signal is given by the standard deviation from the mean signal along the $\phi$ integration.
% Shown in Figure \ref{SIfig:XRD}(d) are the resultant liquid scattering signals
% when propagating the $20\%$ filter thickness uncertainties in both s30970 and s30967. The intensity error of each signal is given by the standard deviation from the mean signal along the $\phi$ integration.
% , divided by the amount of $\phi$ space contributing to the $2\theta$ point.
As we did not record to high-$k$, their absolute signal intensity is not applicable and they are normalized to their broad liquid scattering peak around $45\unit{^{\circ}}$. This process demonstrates the effect filter uncertainties have on the overall shape of the liquid scattering feature.
The total liquid scattering signal uncertainty, as shown in Figure 3(a) of the main paper, therefore covers the region highlighted in gray.

\section{VISAR analysis}{\label{sec:VISAR}}

\begin{figure}[b]
	\centering
	\includegraphics[width=0.995\textwidth]{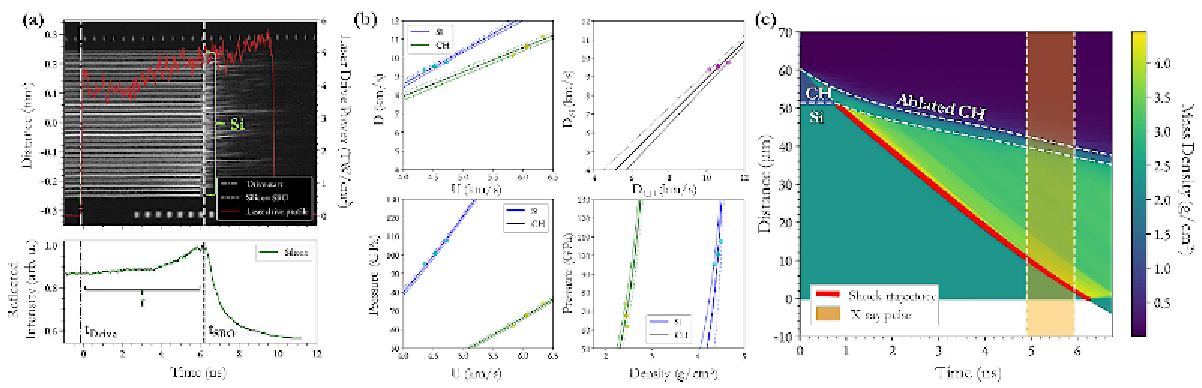}
	\caption{
         \textbf{(a)} Raw VISAR data from s30967. The fringes are reflected from the rear of the $51\,\si{\mu m}$ thick Si sample. Superimposed in red is the laser drive power. The integrated VISAR intensity is projected underneath. The time of shock-breakout, $t_{\mathrm{SBO}}$, is determined as the point at which there is a sharp gradient decline in reflected VISAR signal. 
        \textbf{(b)} Impedance matching CH and Si shock conditions, using equations \ref{SIeqn:D_Si}, \ref{SIeqn:D_CH} and \ref{SIeqn:RHP}, to find (shown in the upper right quadrant) the linear relationship between $D_{\mathrm{CH}}$ and $D_{\mathrm{Si}}$. The statistical errors arising from the CH and Si model uncertainties are shown throughout as fainter green and blue lines, respectively. 
        The appropriate shock speed for each material (highlighted as a magenta diamond) is found by substituting the $D_{\mathrm{CH}}$-$D_{\mathrm{Si}}$ relation into equation \ref{SIeqn:VISARt}.
        The corresponding $1\sigma$ error due to both model and experimental uncertainties are shown as magenta circles.
        This information is carried through the remaining plots to find the corresponding $P$-$\rho$ space for CH and Si.
 	\textbf{(c)} Eulerian HELIOS 1D simulation, produced by scaling the drive laser profile to match the measured $t_{\mathrm{SBO}}$, for s30967. The silicon shock trajectory is shown as a thick red line and the timing of the X-ray laser pulse is highlighted in orange. 
        }
	\label{SIfig:ASBO}
\end{figure}

Impedance matching techniques are used to infer the average shock speed through the silicon sample, $D_{\mathrm{Si}}$.
The total shock time through both the CH ablator and Si sample is $t = t_{\mathrm{SBO}}-t_{\mathrm{drive}}$, where $t_{\mathrm{SBO}}$ and $t_{\mathrm{drive}}$ are the measured shock-breakout (SBO) and laser driver timings.
These timings are listed in Table \ref{SItab:EXP} and an example of the VISAR diagnostic for s30967 is shown in Figure \ref{SIfig:ASBO}(a).
This total time can be related to the shock speeds in the Si and CH via,
\begin{equation}
    \label{SIeqn:VISARt}
    t 
    =
    \frac{h_{\mathrm{Si}}}{D_{\mathrm{Si}}} 
    +
    \frac{h_{\mathrm{CH}}}{D_{\mathrm{CH}}}
    \,,
\end{equation}
where $h$ is the thickness of each material.

As described in the main paper, silicon is opaque to the VISAR laser, meaning a direct measurement of particle velocity, $U$, cannot be obtained on each shot. The particle velocity in the shocked silicon is therefore inferred as \cite{Henderson21},
\begin{equation}
    \label{SIeqn:D_Si}
    D_{\mathrm{Si}}
    =
    10.3(\pm0.1) + 1.8(\pm0.1)\left[U - 4.95\right]\,,
\end{equation}
which is valid for $4<U\,\,\si{(km/s)}<6.5$ and is based on the explosively driven data collected by Pavlovski in Ref. \cite{Pavlovskii68}.
The corresponding linear relationship used for the CH ablator is \cite{Barrios10},
\begin{equation}
    \label{SIeqn:D_CH}
    D_{\mathrm{CH}}
    =
    21.029(\pm0.057) + 1.305(\pm0.015)\left[U - 14.038\right]\,.
\end{equation}
As detailed in Figure \ref{SIfig:ASBO}(b) these equations are used alongside the Rankine-Hugoniot relations, which are derived from the conservation of mass, momentum and energy across a shock front, to infer the post-shock $P$-$\rho$ state,
\begin{equation}
    \label{SIeqn:RHP}
    P
    =
    P_0 + \rho_0U_sU_p
    \,\,\,\,,\,\,\,\,
    \rho
    =
    \frac{\rho_0U_s}{U_s-U_p}\,,
\end{equation}
where $P_0=0\,\si{Mbar}$ and $\rho_0=2.329\,\si{g/cm^3}$ are the pre-shock conditions.
The silicon shock speed, pressure and density for each shock-compressed experiment are listed in Table \ref{SItab:VISARvsHELIOS}.

\begin{table}[t]
    \centering
    \renewcommand{\arraystretch}{1.5}
    \caption{\label{SItab:VISARvsHELIOS} Comparison of the inferred plasma conditions for each shot using Rankine-Hugoniot relations, on the left, and the Si mass-averaged conditions from HELIOS simulations during the scattering event on the right.}
    \begin{ruledtabular}
    \begin{tabular}{l|ccc|ccc}
    \textbf{Shot} & 
    $\mathbf{D_{\mathbf{Si}}}\,\unit{(km/s)}$    & $\mathbf{P}\,\unit{(GPa)}$  & $\boldsymbol{\rho}\,\unit{(g/cm^3)}$
    &$\mathbf{D_{\mathbf{Si}}}\,\unit{(km/s)}$    & $\mathbf{P}\,\unit{(GPa)}$  & $\boldsymbol{\rho}\,\unit{(g/cm^3)}$\\ \hline
    \multicolumn{4}{c|}{\textbf{VISAR Transit Time}} & \multicolumn{3}{c}{\textbf{HELIOS Simulations}}   \\ \hline
    $30964$         & $11.4\pm0.3$          & $146\pm9$     & $4.54\pm0.07$   & $10.9\pm0.2$  & $70\pm7$  & $3.3\pm0.2$ \\
    $30967$         & $9.5\pm0.2$           & $101\pm6$     & $4.43\pm0.08$   & $9.4\pm0.1$   & $48\pm4$     & $3.2\pm0.1$\\
    $33538$         & $11.7\pm0.3$          & $155\pm10$    & $4.56\pm0.07$   & $11.6\pm0.2$  & $81\pm10$       & $3.3\pm0.2$\\
    $33541$         & $12.4\pm0.2$          & $178\pm10$    & $4.60\pm0.07$   & $12.0\pm0.2$  & $82\pm10$       & $3.2\pm0.2$\\ 
    \end{tabular}
    \end{ruledtabular}
\end{table}

\begin{figure}[b]
	\centering
	\includegraphics[width=0.7\textwidth]{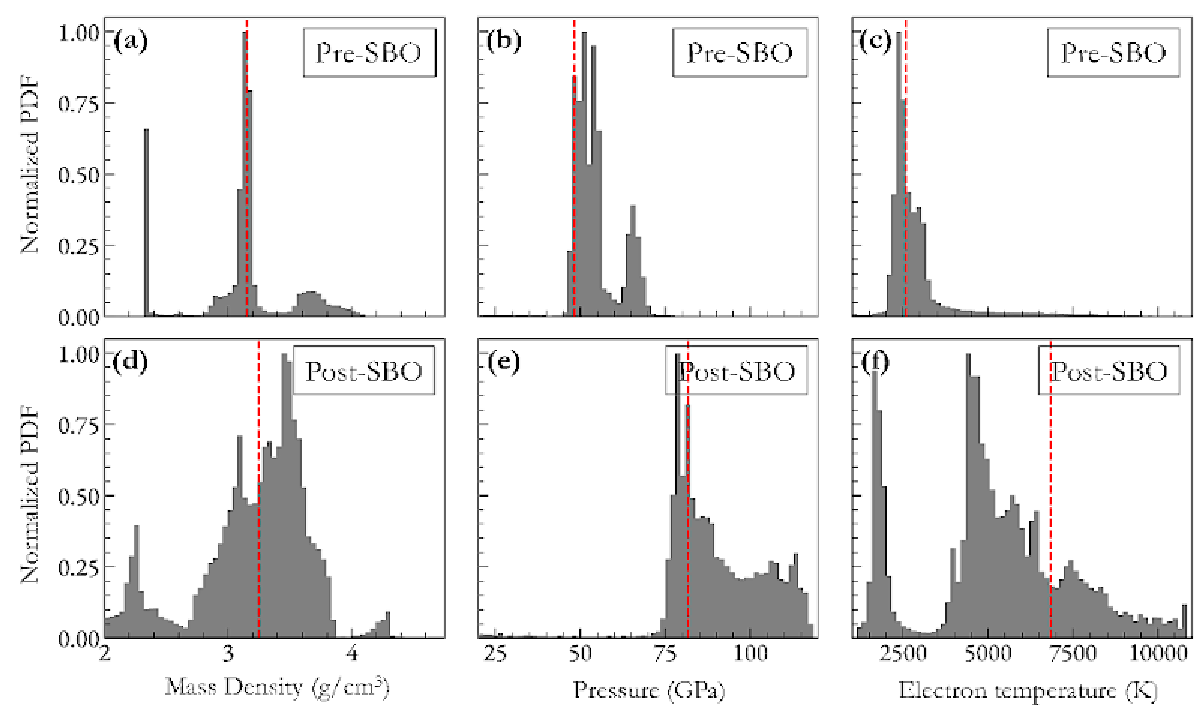}
	\caption{Normalized probability distribution functions (PDFs) of silicon conditions during the X-ray scattering event as predicted by the HELIOS simulations. The vertical red dashed lines are their respective mass-averaged values, $\left<X\right>_m$.
         \textbf{(a)}, \textbf{(b)} and \textbf{(c)} show the density, pressure and $T_e$, respectively, for s30967, taken before shock-breakout.
          \textbf{(d)}, \textbf{(e)} and \textbf{(f)} as above for s33538, taken after shock-breakout.
        }
	\label{SIfig:HELIOS}
\end{figure}

\section{HELIOS Simulations}{\label{sec:HELIOS}}

Using the $t_{\mathrm{SBO}}$ measured with the VISAR diagnostic and by scaling the drive laser profile, HELIOS 1D simulations were produced for each shot using SESAME-EOS \cite{SESAME}. An example of the mass density through the target is shown in Figure \ref{SIfig:ASBO}(c). The simulations emphasize the non-uniformity of the conditions within the silicon during the X-ray scattering event, which is highlighted in orange.
The simulated shock speed through the silicon was determined by tracking the mass density gradient. An example of this shock tracking is shown as a thick red line in Figure \ref{SIfig:ASBO}(c).

The simulated mass-averaged plasma conditions, $\left<X\right>_m$ (where $X$ is a plasma parameter), within the silicon during the X-ray probe were calculated as the mass-weighted average in space and averaged over the probe duration in time \cite{Chapman14}. This is determined as,
\begin{equation}
    \left<X\right>_{m}(t) 
    =
    \frac{\sum_{i} V_{i}(t) \rho_{i}(t) X_{i}(t) }
    { \sum_{i} V_{i}(t) \rho_{i}(t)}  \,,
\end{equation}
\begin{equation}
    \left<X\right>_m 
    \equiv 
    \left<\left<X\right>_{m}(t)\right>_{t} 
    = 
    \frac{\sum_{j_{\mathrm{min}}}^{j_{\mathrm{max}}} \left<X\right>_{m}(t_j)t_j}
    {\Delta t_{\mathrm{xray}}}\,,
\end{equation}
where $V_i(t)$ is the volume in the $i$-th cell at time $t$. The inferred $P$-$\rho$ states for each shot are listed in Table \ref{SItab:VISARvsHELIOS}.
These mass-averaged values are compared to the overall parameter distributions during the X-ray scattering event in Figure \ref{SIfig:HELIOS}. In the pre-SBO data the sharp peaks corresponding to the low density region indicate the presence of ambient silicon.

\section{Markov-Chain Monte Carlo Analysis}{\label{sec:MCMC}}

\begin{figure}[b]
	\centering
	\includegraphics[width=0.995\textwidth]{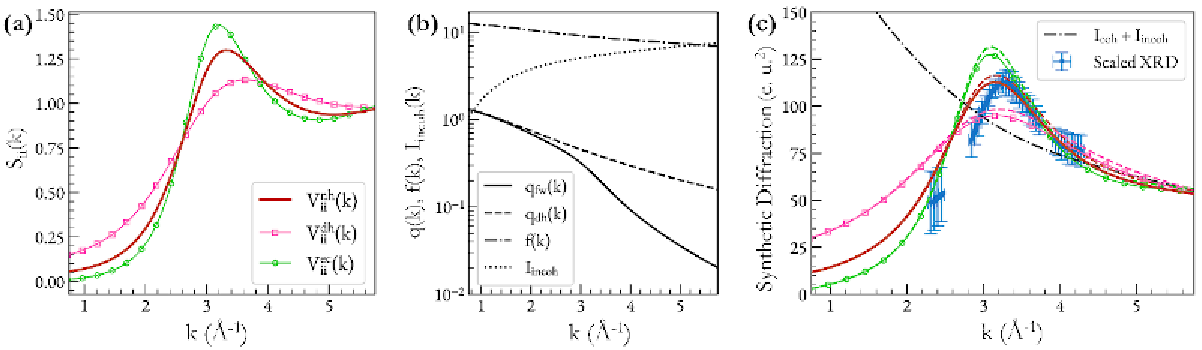}
	\caption{
        \textbf{(a)} Ion-ion structure factors, $S_{ii}(k)$, for each $V_{ii}(k)$ using the mean parameters $\left<\Psi\right>_{\mathrm{NH}}$ as shown in Figure 3(a) of the main paper, ($\rho=4.5\,\si{g/cm^3}$, $Z=1.3$, $T=6800\,\si{K}$).
        \textbf{(b)} Comparison of the tabulated incoherent scattering signal $I_{\mathrm{incoh}}$ \cite{Hubbell75}, the bound electron form factor, $f(k)$, and the screening cloud, $q(k)$, produced with DH and FW models using $\left<\Psi\right>_{\mathrm{NH}}$.
        \textbf{(c)} The synthetic diffraction signals produced with each $V_{ii}(k)$, $q(k)$, $f(k)$ and $I_{\mathrm{incoh}}$, as shown in (a) and (b), compared to the scaled experimental scattering signal. For this representative plot, the scaling parameter, $\Gamma$, from equation \ref{SIeqn:MCMC_scaling} was chosen using the NH $I_{\mathrm{fit}}$ (solid red curve).
        }
	\label{SIfig:MCSS_comp}
\end{figure}

% MCSS models:
% ee lfc = NONE
% ee pol func = NUMERICAL RPA
% ipd model = ION SPHERE
% ei potential  = EFFECTIVE COULOMB

MCMC is a robust method for exploring complex multi-parameter spaces to overcome the challenge of inverse problem instability, which implies that the same measured experimental data can be fitted equally well by very different conditions. 
Given a specific set of parameters, $\Psi(\rho, T, Z)$, the multi-component scattering spectra (MCSS) code \cite{MCSS} is used to produce a theoretical diffraction signal, $I_{\mathrm{fit}}(k)$. As discussed in the main paper, the scaling parameter $\gamma$ for the experimentally measured liquid diffraction signal, $I_{\mathrm{liq}}(k)$, cannot be obtained experimentally. Instead we scale the experimental signal to the simulated fit using,
\begin{equation}
    \label{SIeqn:MCMC_scaling}
    \frac{I_{\mathrm{liq}}(k)}{\gamma} 
    \equiv
    I_{\mathrm{scal}}(k) 
    = 
    I_{\mathrm{liq}}(k)\,\times\,\Gamma\,\,\frac{I_{\mathrm{fit}}^{\mathrm{max}}}{I_{\mathrm{liq}}^{\mathrm{max}}}\,,
\end{equation}
where $\Gamma$ is a free random Gaussian scalar with a standard deviation equal to the noise of the raw data, $I_{\mathrm{fit}}^{\mathrm{max}}$ and $I_{\mathrm{liq}}^{\mathrm{max}}$ are the peak values in the MCSS fit and raw X-ray scattering data, respectively.

The MCMC process then calculates the likelihood of these parameters producing the given scaled X-ray scattering spectrum, $I_{\mathrm{scal}}(k)$, as \cite{Kasim19},
\begin{equation}
    \label{eqn:bayesian}
    P(\Psi |I_{\mathrm{scal}}(k))
    =
    \frac{P\left(I_{\mathrm{scal}}(k)|\Psi\right)P(\Psi)}{P(I_{\mathrm{scal}}(k))}\,,
\end{equation}
where $P(\Psi)$ is the prior distribution of possible parameters, $P(I_{\mathrm{scal}}(k))$ is the marginal likelihood of the observed data over all possible parameters and the forward model likelihood,  $P(I_{\mathrm{scal}}(k)|\Psi)$, is as described in the main paper.

\begin{figure}[b]
	\centering
	\includegraphics[width=0.995\textwidth]{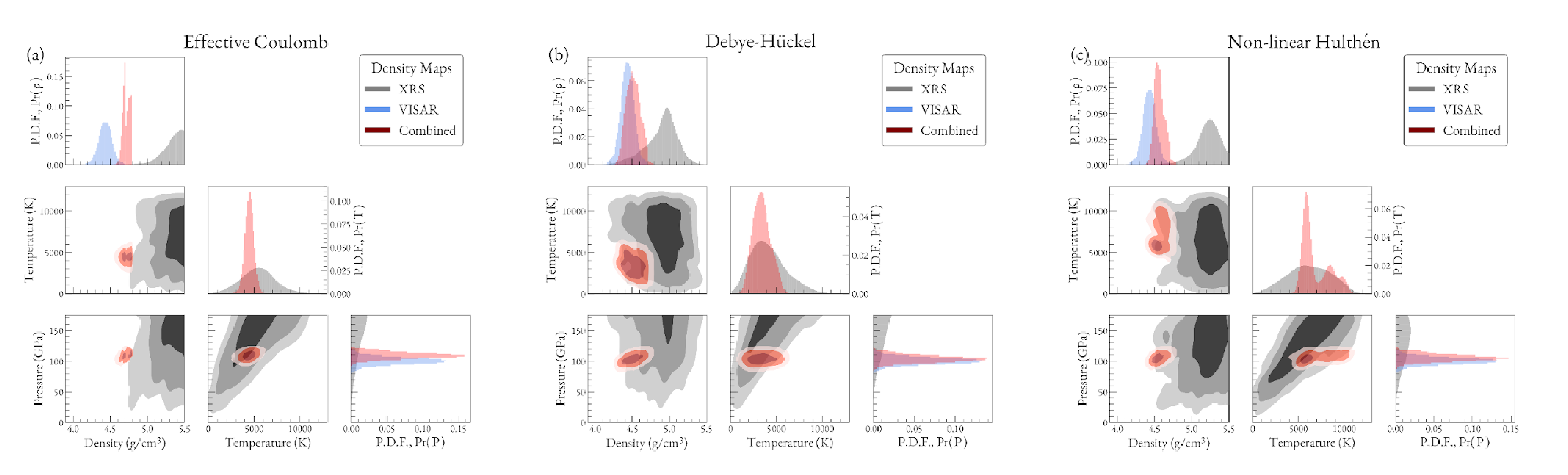}
	\caption{
        Probability density functions for the liquid silicon density, pressure and temperature state using the Debye-H\"uckel \textbf{(a)}, effective Coulomb \textbf{(b)} and non-linear Hulth\'en \textbf{(c)} $V_{ii}(k)$ models.
        As in Figure 3 of the main paper, the lower quadrant plots compare the $1$, $2$ and $3\sigma$ parameter correlations for the MCMC converged X-ray scattering analysis (gray heat maps) and the combined density functions (red heat maps).
        The diagonal histograms show the probability densities for each parameter. The VISAR distributions are added in blue for density and pressure.
                }
	\label{SIfig:MCMC}
\end{figure}

Model sensitivities for producing synthetic X-ray scattering signals using MCSS are shown in Figure \ref{SIfig:MCSS_comp}. This demonstrates how the dominant contributor to the synthetic scattering signal is the chosen ion-ion interaction potential, $V_{ii}(k)$. The screening models, $q(k)$, cannot be differentiated within the measured experimental error. The potentials for each $V_{ii}(k)$ are given by,
\begin{equation}
    \label{SIeqn:ViiEC}
    V_{ii}^{ec}(k)
    =
    \frac{\mathcal{K}}{k^2}\,,
\end{equation}
where $\mathcal{K}=-4\pi Z^2e^2/3k_BT_e$,
\begin{equation}
    \label{SIeqn:ViiDH}
    V_{ii}^{dh}(k)
    =
    \frac{\mathcal{K}}{\kappa_e^2+k^2}\,,
\end{equation}
where $\kappa_e$ is the inverse screening length, and
\begin{equation}
    \label{SIeqn:ViiNH}
    V_{ii}^{nh}(k)
    =
    \mathcal{K}\frac{i}{4k\kappa_e}
    \left[{\psi^{(1)}\left(1+i\frac{k}{2\kappa_e}\right)
    -
    \psi^{(1)}\left(1-i\frac{k}{2\kappa_e}\right)}\,,
    \right]
\end{equation}
where $\psi^{(1)}(x)$ is a trigamma function.
The inverse screening length is taken as,
\begin{equation}
    \label{SIeqn:screening}
    \kappa_e
    =
    \sqrt{\frac{n_e e^2}{k_B T_e \epsilon_0}\frac{\mathcal{F}_{-1/2}(\eta_{e})}{\mathcal{F}_{1/2}(\eta_{e})}}\,,
\end{equation}
where $\mathcal{F}_j$ denotes the complete Fermi-Dirac integral of order $j$ and $\eta_e$ is the dimensionless chemical potential of the free electrons,
\begin{equation}
    \label{SIeqn:chemical}
    \eta_e
    =
    \mathcal{F}_{1/2}^{-1}\left[
    \frac{n_e\hbar^3}{2}\left(\frac{2\pi}{m_ek_BT_e}\right)^{3/2}\right]\,.
\end{equation}
The converged shock-compressed silicon conditions using the effective Coulomb, Debye-H\"uckel and non-linear Hulth\'en models are shown in gray in Figure \ref{SIfig:MCMC}.

From the convergence parameters, the total plasma pressure is determined using the `two-fluid' model described by Vorberger \textit{et al.}\;\cite{Vorberger13, Ebeling20},
\begin{align}
    \label{P_tot}
    P
    = &\,
    \left(1 + \langle{Z_{i}}\rangle\dfrac{\mathcal{F}_{3/2}(\eta_{e})}{n_{e}\Lambda_{e}^{3}/2}\right)n_{i}k_{\text{B}}T
    -
    n_{i}^{2}\left.
    \dfrac{\partial f_{e}^{\text{XC}}}{\partial n_{i}}
    \right|_{T}
    \nonumber\\
    & -
    \dfrac{n_{i}^{2}}{4\pi^2}
    \dfrac{\partial }{\partial n_{i}}
    \left.\!\!\left(
    \int_{0}^{\infty}\text{d}k\,k^{2}\,
    \chi_{ee}(k,0)\!\left[V_{ei}^{\text{eff}}(k)\right]^{2}
    \right)
    \right|_{T}
    \nonumber\\
    & -
    2\pi n_{i}^{2}
    \int_{0}^{\infty}\text{d}r\,r^{2}
    \left[g_{ii}(r; V_{ii}^{\text{eff}}(r)) - 1\right]
    \mathcal{V}_{ii}^{\text{eff}, P}(r)
    \,,
\end{align}
in which the effective ion-ion interaction potential for the evaluation of the excess ion pressure is defined as
\begin{align}
    \label{eff_pot}
    \mathcal{V}_{ii}^{\text{eff}, P}(r)
    = &\,
    \left(\dfrac{r}{3}\dfrac{\partial}{\partial r} - n_{i}\dfrac{\partial}{\partial n_{i}}\right)
    V_{ii}^{\text{eff}}(r)
    \,.
\end{align}
In Eq.\;\eqref{P_tot}, $\Lambda_e$ is the thermal de Broglie wavelength of the free electrons. The first term is the sum of the ideal gas contributions from the ions and free electrons. In the second term, $f_{e}^{\text{XC}}$ is the exchange-correlation contribution to the free energy (per particle) of the interacting electron gas \cite{Groth17}. The third term gives the contribution arising from electron-ion interactions in linear response, where the static electron density response function, $\chi_{ee}(k)$, is taken in the random phase approximation and the local field correction, $G_{ee}(k)$, is determined using the effective static approximation \cite{Dornheim21}.
The fourth term gives the excess pressure due to ionic correlations, where $g_{ii}(r; V_{ii}^{\text{eff}}(r))$ is the pair distribution function of the ions \cite{Hansen} evaluated with a consistent effective ion-ion potential. 

\begin{table}[t]
    \centering
    \renewcommand{\arraystretch}{1.5}
    \caption{\label{SItab:MCMCvsCOMB} Comparison of the liquid silicon conditions within $1\sigma$ for the X-ray scattering MCMC convergence and the combined VISAR state using each ion-ion interaction potential.}
    \begin{ruledtabular}
    \begin{tabular}{l|cccc}
    \textbf{$\mathbf{V_{ii}(k)}$} & 
    $\boldsymbol{\rho}\,\unit{(g/cm^3)}$    & $\mathbf{P}\,\unit{(GPa)}$  & $\mathbf{T}\,\unit{(K)}$ &$\mathbf{\Bar{Z}}$\\ \hline
    \multicolumn{5}{c}{\textbf{MCMC Convergence}}  \\ \hline
    EC  & $5.5\pm0.2$   & $200\pm60$    & $6600\pm2900$    & $1.3\pm0.3$ \\
    DH  & $4.9\pm0.2$   & $250\pm90$    & $6900\pm3000$    & $3\pm1$   \\
    NH  & $5.2\pm0.2$   & $140\pm40$    & $6500\pm3000$    & $1.5\pm0.4$   \\ \hline
    \multicolumn{5}{c}{\textbf{MCMC and VISAR Combined}}  \\ \hline
    EC  & $4.72\pm0.05$   & $110\pm5$    & $4500\pm500$     & $1.01\pm0.03$  \\
    DH  & $4.51\pm0.09$   & $104\pm6$    & $3400\pm1000$    & $1.5\pm0.1$  \\
    NH  & $4.57\pm0.07$   & $106\pm6$    & $6900\pm1500$    & $1.46\pm0.06$  \\
    \end{tabular}
    \end{ruledtabular}
\end{table}
% \color{red}
% The numerical techniques used to evaluate \eqref{P_tot} are discussed in the Supplemental Material.
% \color{black}

\section{Combining VISAR and MCMC analysis}{\label{sec:Combining}}

As shown in Figure 3 of the main paper, given the VISAR and MCMC $P$-$\rho$ probability density functions, their subsequent joint probability is defined as,
\begin{equation}
    \label{SIeqn:jointP}
    \mathrm{Pr}_j(\rho, P)
    =
    \frac{\mathrm{Pr}_m(\rho, P)\times\mathrm{Pr}_v(\rho, P)}{\sum_{\rho, P}\left[\mathrm{Pr}_m(\rho, P)\times\mathrm{Pr}_v(\rho, P)\right]}\,.
\end{equation}
These 2D combined density functions are shown as red heat maps in the lower left plots in Figure \ref{SIfig:MCMC}.
This formalism can be extended into multi-parameter dimensions as the X-ray scattering analysis has dependencies beyond density and pressure.
However, as VISAR provides no direct measurement of temperature, we must define its 3D density function such that $\mathrm{Pr}_v(\rho, P, T_i)\equiv\mathrm{Pr}_v(\rho, P, T_j)$, where $i(\neq j)$ describes a position along the temperature axis.
By using equation \ref{SIeqn:jointP}, the 2D phase space for pressure and temperature can subsequently be found as,
\begin{equation}
    \label{SIeqn:2Dfrom3D}
    \mathrm{Pr}_j(P, T)
    =
    \frac{\sum_{\rho}\mathrm{Pr}_j(\rho, P, T)}{\sum_{P, T}\left[\sum_{\rho}\mathrm{Pr}_j(\rho, P, T)\right]}\,.
\end{equation}
This process can be repeated in the ionization space.

The $1\sigma$ errors of each parameter's joint probability density function, i.e. $\mathrm{Pr}_j(\rho)$, are given in Table \ref{SItab:MCMCvsCOMB}. The phase diagrams in Figure 4 of the main paper detail the $1\sigma$ errors from $\mathrm{Pr}_j(\rho, P)$ and $\mathrm{Pr}_j(P, T)$, shown as the dark red contours in the lower left and central density maps in Figure \ref{SIfig:MCMC}.

\bibliographystyle{apsrev}
\bibliography{supplement}